\documentclass[a4paper,twocolumn,preprintnumbers,citeautoscript,newabstract,aps,superscriptaddress, %floatfix%,nofootinbib
]{revtex4} %floatfix,floats
\usepackage{amsmath,amsfonts,amssymb,graphics}
\usepackage[normalem]{ulem}
\usepackage{graphicx}
\usepackage{units}
\usepackage{mathrsfs}
\usepackage{bbm}
\usepackage{pifont}
\usepackage{braket}
\usepackage{units}
\usepackage{footmisc}
\usepackage[dvipsnames]{xcolor}
\usepackage{mathtools}
\usepackage{xspace}
\usepackage{booktabs}

\usepackage[bookmarks=false,hyperfootnotes=false]{hyperref}%\usepackage{hyperref}

\definecolor{nicered}{rgb}{0.5,.0,.0}
\definecolor{darkblue}{rgb}{0,.1,.9}
\definecolor{lightblue}{rgb}{0,.1,.6}
 \definecolor{darkgreen}{rgb}{0.0,0.2,0.0}
\hypersetup{colorlinks, citecolor=darkgreen ,linkcolor=darkblue, urlcolor=lightblue}

\newcommand{\Figref}[1]{Fig.~\ref{#1}}
\newcommand{\Tabref}[1]{Tab.~\ref{#1}}

\newcommand{\keV}{\ensuremath{\,\mathrm{keV}}\xspace}
\newcommand{\MeV}{\ensuremath{\,\mathrm{MeV}}\xspace}
\newcommand{\GeV}{\ensuremath{\,\mathrm{GeV}}\xspace}
\newcommand{\eV}{\ensuremath{\,\mathrm{eV}}\xspace}
\newcommand*{\rep}[2][]{\ensuremath{{\boldsymbol{#2}#1}}}

\newcommand{\I}{\mathrm{i}}

\newcommand{\SU}[1]{\ensuremath{\mathrm{SU}(#1)}}
\newcommand{\U}[1]{\ensuremath{\mathrm{U}(#1)}}
\newcommand{\Up}{\ensuremath{\mathrm{U}(1)_{\mathrm{X}}}\xspace}
\newcommand{\N}[1]{\ensuremath{N_{#1}}}
\newcommand{\Nb}[1]{\ensuremath{\overline{N}_{#1}}}
\newcommand{\vH}{\ensuremath{v_h}\xspace}
\newcommand{\vP}{\ensuremath{v_{\phi}}\xspace}
\newcommand{\vS}{\ensuremath{v_s}\xspace}
\newcommand{\vbar}{\ensuremath{\bar{v}}\xspace}
\newcommand{\Zp}{\ensuremath{Z'}\xspace}
\newcommand{\hS}{h_{S}\xspace}
\newcommand{\lI}{\ensuremath{\lambda_3}}
\newcommand{\lII}{\ensuremath{\lambda_4}}

\newcommand{\eM}{\ensuremath{\varepsilon_m}\xspace}
\newcommand{\mZp}{\ensuremath{m_{\Zp}}}
\newcommand{\Mpl}{\ensuremath{M_{\mathrm{Pl}}}}
\newcommand{\Geff}{\ensuremath{G^{4\nu}_{\mathrm{eff}}}}
\newcommand{\geff}{\ensuremath{g_{\mathrm{eff}}}}
\newcommand{\dNeff}{\ensuremath{\Delta N_{\mathrm{eff}}}\xspace}
\newcommand{\Neff}{\ensuremath{N_{\mathrm{eff}}}\xspace}
\newcommand{\myvec}[1]{\ensuremath{\begin{pmatrix}#1\end{pmatrix}}}

\definecolor{darkgreen}{rgb}{0.0, 0.6, 0.2}

\allowdisplaybreaks[1]

\begin{document}

%\preprint{preprint#\\}

\title{\textbf{\boldmath \Large The Hubble tension and a renormalizable model of \\[0.1cm] gauged neutrino self-interactions \unboldmath}}

\author{Maximilian Berbig}
\email[]{berbig@physik.uni-bonn.de}
\affiliation{Bethe Center for Theoretical Physics und Physikalisches 
Institut der Universit\"at Bonn, \\ Nussallee 12, 53115 Bonn, Germany}
\author{Sudip Jana}
\email[]{sudip.jana@mpi-hd.mpg.de}
\affiliation{Max-Planck-Institut f\"ur Kernphysik, Saupfercheckweg 1, 69117 Heidelberg, Germany}
\author{Andreas Trautner}
\email[]{trautner@mpi-hd.mpg.de}
\affiliation{Max-Planck-Institut f\"ur Kernphysik, Saupfercheckweg 1, 69117 Heidelberg, Germany}
%
%\date{\today}

\begin{abstract}
We present a simple extension of the Standard Model that leads to renormalizable long-range vector-mediated neutrino self-interactions.
This model can resolve the Hubble tension by delaying the onset of neutrino free-streaming during recombination,
without conflicting with other measurements. 
The extended gauge, scalar and neutrino sectors lead to observable signatures, including invisible Higgs and $Z$ decays, thereby
relating the Hubble tension to precision measurements at the LHC and future colliders. 
The model has a new neutrinophilic gauge boson with $\mZp\sim\mathcal{O}(10~\eV)$ and charged Higgses at a few $100~\GeV$.
It requires hidden neutrinos with active-hidden mixing angles larger than $5\times10^{-4}$
and masses in the range $1\div300\eV$, which could also play a role for short baseline neutrino oscillation anomalies.
\end{abstract}

\maketitle
\enlargethispage{1.5cm}
{\textbf{\textit{Introduction.---}}}%
There is convincing evidence that neutrinos played a substantial role during the 
epoque of big bang nucleosynthesis (BBN) at $T\sim\mathrm{MeV}$, 
closely monitored by early element abundances. 
The lowest temperature scale indirectly probed for neutrinos is $T\sim\mathrm{eV}$, 
where observations of the cosmic microwave background (CMB) 
fit well to a history of our universe that does not only
comply with the cosmological standard model ($\Lambda$CDM), 
but also with the expectation of the Standard Model of
particle physics (SM), including exactly three generations of neutrinos.

However, evidence is accumulating \textit{not only} for a discrepancy between local measurements 
of today's Hubble rate $H_0$ \cite{Riess:2016jrr,Riess:2019cxk, Freedman:2019jwv,Birrer:2018vtm,Wong:2019kwg} 
and therelike global determinations based on $\Lambda$CDM together with CMB \cite{Aghanim:2018eyx},
baryonic acoustic oscillations (BAO) and large scale structure (LSS) datasets 
\cite{Kohlinger:2017sxk,Joudaki:2017zdt,Joudaki:2019pmv,Alam:2016hwk,Philcox:2020vvt,Troxel:2017xyo,Abbott:2017smn,Abbott:2017wcz,Hikage:2018qbn,Abdullah:2020qmm},
\textit{but also} for an increasing tension in other parameters \cite{DiValentino:2019qzk,DiValentino:2020hov}.
Resolving these discrepancies might require a modification of $\Lambda$CDM, preferentially, perhaps, shortly before the era of recombination~\cite{Schoneberg:2019wmt,Knox:2019rjx}.
Many new physics (NP) scenarios have been discussed, see~\cite{Knox:2019rjx, Escudero:2019gvw, Sakstein:2019fmf, Verde:2019ivm}. 
Naturally, any consistent modification of $\Lambda$CDM must be in compliance with a consistent modification of the Standard Model of particle physics (SM).

The positive correlation of $H_0$ and $\Neff$ with the amplitude of the matter power spectrum 
$\sigma_8$, as observed in CMB data \cite{Aghanim:2018eyx}, 
prohibits a resolution of the $H_0$ tension simply by increasing $\Neff$ alone (LSS prefers low $\sigma_8$).
However, a delay in the onset of neutrino free-streaming during recombination could achieve both: 
breaking the positive correlation of $H_0$ and $\sigma_8$, while solving the Hubble tension (HT) at the cost 
of increasing \dNeff during recombination \cite{Cyr-Racine:2013jua,Archidiacono:2013dua,Lancaster:2017ksf,Oldengott:2017fhy,Kreisch:2019yzn,Park:2019ibn}.
Taking into account an effective four-neutrino interaction  $G_{\mathrm{eff}}^{4\nu}(\bar{\nu}\nu)(\bar{\nu}\nu)$ 
a good, bi-modal fit to CMB data is obtained with \cite{Kreisch:2019yzn, Park:2019ibn}
\begin{equation}\label{eq:G4nu}
 G_{\mathrm{eff}}^{4\nu}\equiv\frac{\geff^2}{ m_{\Zp}^2}\approx
 \begin{cases}
  \left(5\,\MeV\right)^{-2}  & \text{(SI),\quad or}\\
  \left(100\,\MeV\right)^{-2} & \text{(WI)}\;.
\end{cases}
\end{equation}
The weakly interacting mode (WI) should be interpreted
as an upper limit on $G_{\mathrm{eff}}^{4\nu}$ such that cosmological parameters stay 
close to $\Lambda$CDM \cite{Cyr-Racine:2013jua,Archidiacono:2013dua}, without resolving above tensions.
Therefore, we focus on the strongly interacting mode (SI), which considerably alters cosmology to resolve the tensions in $H_0$ and $\sigma_8$
while being consistent with local astronomical observations \cite{Kreisch:2019yzn, Park:2019ibn}.

\begin{figure}[t]
\centering\includegraphics[width=1.0\linewidth]{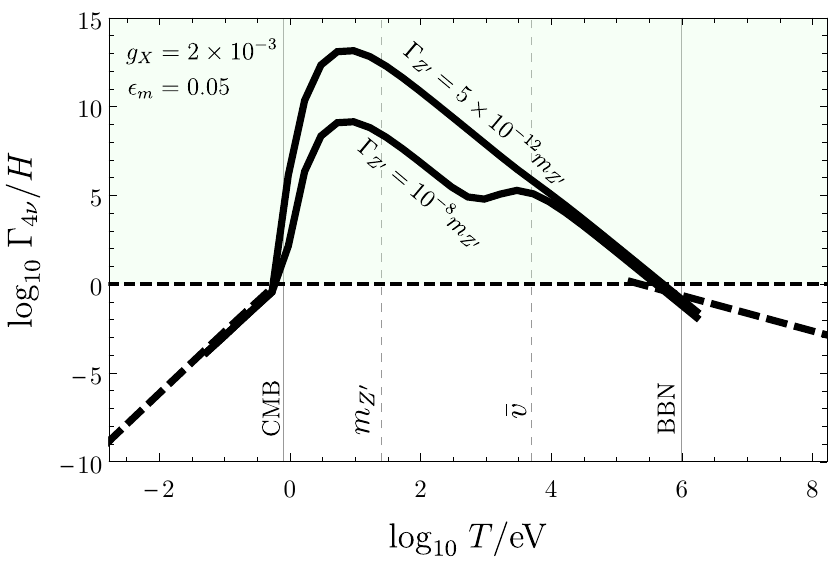}
\caption{\label{fig:GamVsH}
Thermally averaged four-neutrino interaction rate relative to the Hubble rate as a function of Temperature
for $\mZp=25\eV$ and two different values of the $Z'$ width.}
\end{figure}

A valid alternative to heavy new physics would be to generate 
$G_{\mathrm{eff}}^{4\nu}$ by the exchange of a light mediator.
However, it is basically excluded to have very light mediators that recouple during recombination~\cite{Forastieri:2015paa,Forastieri:2019cuf}
and the same is true for light mediators which are thermalized during BBN~\cite{Fields:2019pfx}.
Nontheless, six orders of magnitude in temperature between BBN and CMB are enough to establish a mass scale, 
say after a phase transition, and subsequently integrate it out 
to obtain a decoupling behavior of neutrinos during CMB resembling~\eqref{eq:G4nu}.
In this way, neutrinos \textit{re}couple by the new interactions only after BBN, 
and fall out of equilibrium during recombination, see \Figref{fig:GamVsH}.

Here we provide arguably the simplest renormalizable and phenomenologically viable
extension of the SM that leads to vector mediated four-neutrino interactions of above characteristic.
We first outline the parameter space suitable to address the HT, then present our 
model and discuss its phenomenology.

\medskip
{\textbf{\textit{Relevant parameter region.---}}}%
The effective four-neutrino interaction strength in our model is
\begin{equation}\label{eq:G4nueff}
 G_{\mathrm{eff}}^{4\nu}\equiv\frac{g^2_{\mathrm{eff}}}{ m_{\Zp}^2}\equiv\frac{g_X^2\,\eM^4}{ m_{\Zp}^2}\;,
\end{equation}
where $g_X$ is the gauge coupling of a new \Up symmetry, 
$\eM\ll1$ a mixing between active and hidden (\Up charged) neutrinos,
and $m_{\Zp}$ the mass of the new gauge boson after \Up breaking.
Equating the resulting thermally averaged interaction rate with the Hubble rate,
${G_{\mathrm{eff}}^{4\nu}}^2T^5\approx T^2/\Mpl$, confirms $T_{\mathrm{dec.}}\approx 0.5\,\mathrm{eV}$.

On the other hand, for a range of temperatures \mbox{$T\gg m_{\Zp}$}, while $\eM$ is relevant, 
the new gauge boson will be effectively massless, inducing a long-range four-neutrino interaction with thermally averaged rate
\mbox{$\Gamma\sim\geff^4\,T$}. Requiring this interaction not to thermalize neutrinos prior to BBN, but before recombination, demands
\begin{equation}\label{eq:geffWindow}
 2\times10^{-7}\lesssim g_X\,\eM^2\lesssim 5\times10^{-6}\;.
\end{equation}
Knowing $\geff$ and $G_{\mathrm{eff}}^{4\nu}$ we can compute
\begin{equation}\label{eq:mZp}
 1\lesssim \mZp \lesssim 25\,\mathrm{eV}\quad \text{(SI)}.
\end{equation}
Parametrizing $\mZp=g_X\,\vbar$
we can furthermore constrain the effective \Up-breaking vacuum expectation value~(VEV) 
\begin{equation}\label{eq:vbar}
 \vbar:=\frac{\mZp}{g_X}\approx \eM^2\times 5\,\mathrm{MeV}\quad \text{(SI).}
\end{equation} 
\Figref{fig:GamVsH} illustrates this particular re- and decoupling behavior, 
computed exactly within our model. 
Eq.~\eqref{eq:vbar} implies a hierarchy between the relevant scales in the model of
\begin{equation}\label{eq:xi}
 \xi:=\bar{v}/\vH \approx \eM^2\times2\times10^{-5}\quad \text{(SI)}\;,
\end{equation}
where $\vH=246\,\GeV$ is the SM Higgs VEV. 

\medskip
{\textbf{\textit{Model.---}}}% 
\renewcommand{\arraystretch}{1.5}
\begin{table}[t]
\begin{center}
\begin{tabular}{lccccc}
  \toprule[1pt]
  Field & $\Phi$ & $\N1$ & $\N2$ & $S$ & $X_\mu$ \\
  \hline 
  Lorentz & S & RH & RH & S & V \\
  $\SU2_{\mathrm{L}}\times\U1_{\mathrm{Y}}$ & $(\rep{2},-\frac12)$ & $\emptyset$ & $\emptyset$ & $\emptyset$ & $\emptyset$ \\
  \Up & $+1$ & $+1$ & $-1$ & $+1$ & $0$ \\
  $\U1_{\mathrm{L}}$ & $0$ & $+1$ & $-1$ & $0$ & $0$ \\
  \bottomrule[1pt]
  \end{tabular}
  \end{center}
  \caption{\label{tab:model} 
  New fields and their charges under Lorentz, SM gauge, new $\Up$ gauge symmetry 
  as well as under global Lepton number (S=Scalar, RH=right-handed Weyl fermion, V=vector).}
\end{table}
Next to the new \Up gauge symmetry we introduce a pair of SM-neutral chiral fermions $\N{1,2}$
and two new scalars $\Phi,S$ with charges shown in \Tabref{tab:model}~\footnote{%
A model similar to ours, albeit in a different parameter region has been investigated in \cite{Farzan:2016wym,Farzan:2017xzy,Denton:2018dqq}.
However, baryons are neutral under our \Up, implying the absence of large non-standard neutrino-matter interactions.}.
New interactions for SM-leptons are
\begin{equation}\label{eq:Lint}
\mathcal{L}_\mathrm{\mathrm{new}} = -y\,\bar{L}\,\tilde{\Phi}\,\N1-M\,\N1\,\N2 + \mathrm{h.c.},
\end{equation}
where $\tilde{\Phi}:=\mathrm{i}\sigma_2\Phi^*$, $y$ is a Yukawa coupling, and $M$ has mass-dimension one.
We treat the one generation case here, but consider
three generations of SM leptons and multiple generations of hidden fermions below.

We consider the most general scalar potential consistent with
all symmetries,
\begin{equation}
 V=V_H+V_\Phi+V_S+V_{H\Phi}+V_{HS}+V_{\Phi S}+V_3,
\end{equation}
with
\begin{align}
 V_{\Sigma}&:=\mu_\Sigma^2\, \Sigma^\dagger \Sigma + \lambda_\Sigma\, \left(\Sigma^\dagger \Sigma\right)^2 \quad\!\! (\Sigma=H,\Phi,S)\;, \\
 V_{H\Phi}&:=\lI\,\left(H^\dagger H\right)\left(\Phi^\dagger \Phi\right) + \lII\,\left(H^\dagger \Phi\right)\left(\Phi^\dagger H\right), \raisetag{12pt}\\ 
 V_{DS}&:=\lambda_{DS}\,\left(D^\dagger D\right)\left(S^* S\right)\qquad\quad (D=H,\Phi)\;,\\
 V_3&:=-\sqrt{2}\,\mu\,\left(H^\dagger \Phi\right)S^* + \mathrm{h.c.}\;.
\end{align}
The scalars are decomposed as~\mbox{$S=\frac{1}{\sqrt{2}}\left(s + \I a_s\right)$},
\begin{align}
 &H=\myvec{h^+ \\ \frac{1}{\sqrt{2}}\left(h + \I a_h  \right)},\;\text{and}\;\; 
 \Phi=\myvec{\phi^+ \\ \frac{1}{\sqrt{2}}\left(\phi + \I a_{\phi} \right)}.
\end{align}
We choose parameters such that all neutral scalars obtain VEVs \mbox{$v_\sigma:=\langle \sigma\rangle$} for \mbox{$\sigma=h,\phi,s$},
and assume CP conservation in the scalar sector. $\vH$ spontaneously breaks EW-symmetry, $\vS$ breaks \Up, and $\vP$ breaks both. 
Fixing the HT requires $\vH\gg\vS,\vP$, cf.\ \eqref{eq:xi}, and we expand all of our expressions to leading order in that hierarchy.

The photon is exactly the same massless combination of EW bosons as in the SM, mixed by the electroweak angle $c_W:=m_W/m_Z$~\footnote{%
We use abbreviations $\sin\theta_i\equiv s_i$, $\cos\theta_i\equiv c_i$, and $\tan\theta_i\equiv t_i$  for all angles $\theta_i$ in this work.}.
By contrast, the very SM-like $Z$ boson contains a miniscule admixture of the new gauge boson $X$,
\begin{align}
Z_\mu &= c_X\left(c_W\,W^3_\mu - s_W\,B_\mu\right) + s_X\,X_\mu\;,
\end{align}
with an angle~\footnote{%
$s_X$ can be modified by gauge-kinetic mixing,
we comment on this below Eq.~\eqref{eq:NSI}.}
\begin{equation}\label{eq:sinX}
 s_X\approx-2\,c_W\frac{g_X}{g_2}\left(\frac{\vP}{\vH}\right)^2\lll1\quad\text{and}\quad c_X\approx1\;.
\end{equation}
Masses of the neutral gauge bosons up to $\mathcal{O}(\xi^2)$ are
\begin{equation}
 m_Z\approx\frac{g_2\,\vH}{2c_W}\quad\text{and}\quad m_{\Zp}\approx g_X\,\vbar:=g_X\sqrt{\vP^2+\vS^2}\;.
\end{equation}
The neutrino mass matrix in the gauge basis $\left(\nu,\Nb1,\Nb2\right)$ is
\begin{equation}\label{eq:Mnu}
M_{\nu} =
\begin{pmatrix}
0 & -y \vP/\sqrt{2} & 0 \\
-y \vP/\sqrt{2} & 0 & M\\ 
0 & M & 0\\
\end{pmatrix}.
\end{equation}
Upon $13$--rotating by an angle $\eM$,
\begin{equation}\label{eq:eM}
\tan \eM:=(y\vP)/(\sqrt{2}M)\;,
\end{equation}
$M_{\nu}$ reveals an \textit{exact} zero eigenvalue, corresponding to approximately massless active neutrinos,
and a Dirac neutrino $N$ with $M_N:=\sqrt{M^2+ y^2 \vP^2/2}$.
The massless active neutrinos mix with $\Nb2$ proportional to $s_{\eM}$
generating
\eqref{eq:G4nueff}. Defining $\tan \gamma:=\vP/\vS,$
one can show that
\begin{equation}\label{eq:Mbound}
 M=(y/\sqrt{2})\,\eM\,s_\gamma\,(\Geff)^{-1/2}~\ll~5\,\MeV\;.
\end{equation}
Owing to constraints discussed below the parameter range 
one should have in mind is \mbox{$2\times10^{-5}\lesssim y \lesssim 6\times 10^{-3}$},
$\eM\lesssim 0.05$ and $s_\gamma\lesssim 0.2$.
$M_N\approx M$ then turns out to be 
in the range $1\div300\,\mathrm{eV}$.

Non-zero active neutrino masses are required by neutrino oscillation phenomenology, 
and also for a successful resolution of the Hubble tension
with self-interacting neutrinos~\cite{Kreisch:2019yzn, Park:2019ibn}.
The mass generation for active neutrinos \mbox{$m_\nu\ll y\vP$}
can be realized as a small perturbation to the diagonalization of
$M_\nu$.
In particular, our mechanism is compatible with an 
effective Majorana mass in $\left[M_\nu\right]_{11}$,
and, therefore, with any type of mass generation mechanism
that gives rise to the Weinberg operator \cite{Weinberg:1979sa}.
Another minimal possibility in the present model would be to populate 
$\left[M_\nu\right]_{33}$ like in the inverse seesaw mechanism 
\cite{Mohapatra:1986su,Bertuzzo:2018ftf, Dev:2012sg}.
Also Dirac masses are possible but require additional fermions.
This shows that, ultimately, any of the commonly considered neutrino 
mass generation mechanisms is compatible with our model.

\begin{figure}[t]
\centering\includegraphics[width=1.0\linewidth]{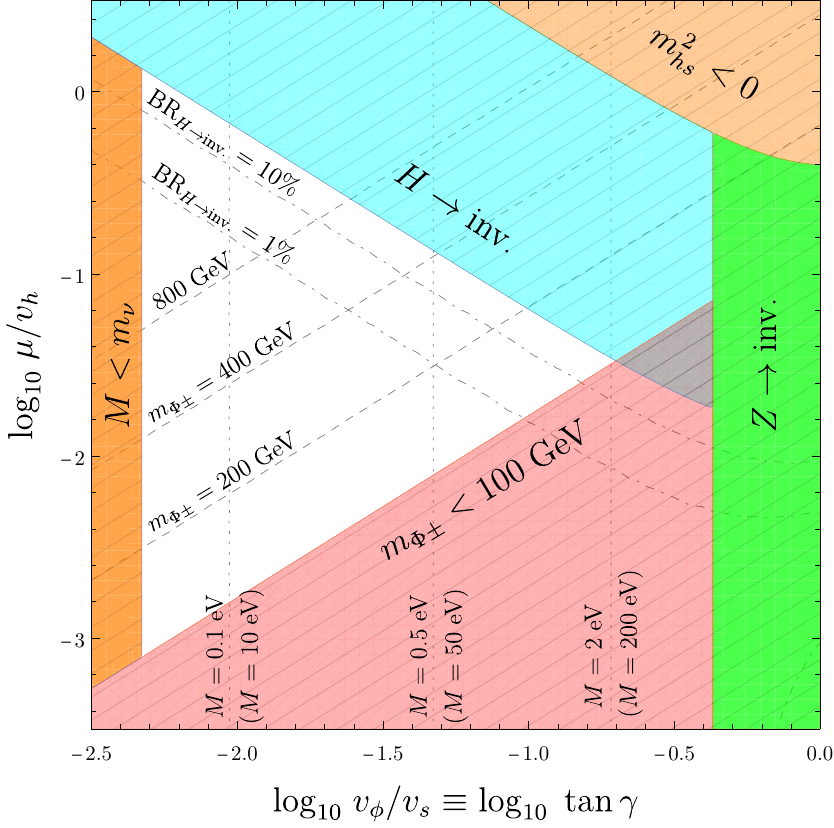}
\caption{\label{fig:mugamma}
Allowed region (blank) in the \mbox{$\tan\gamma$--$(\mu/\vH)$} plane.
The region is independent of any other free parameters 
as long as $\lambda_{HS}, \lambda_{34}\ll\mathcal{O}(10^{-2})$
(for definiteness, we have chosen scalar potential parameters as 
 $\lambda_{HS}=0.001$,
 $\lambda_3=0.002$,
 $\lambda_4=0.003$,
 $\lambda_{\Phi}=0.3$,
 $\lambda_S=0.4$,
 $\lambda_{\Phi S}=0.5$).
The Hubble tension can be resolved in the entire allowed region.
We show equilines of the corresponding hidden neutrino mass $M$ for benchmark 
parameters $\eM=0.05$, $g_X=2\times10^{-3}$, and $y=6\times10^{-5}(6\times10^{-3})$.
In the orange region $M<m_\nu$ (for $y=6\times10^{-5}$) which is inconsistent with our assumptions.}
\end{figure}

\medskip
{\textbf{\textit{Phenomenology.---}}}%
The scalar sector of the model corresponds to 2HDM+scalar singlet. 
However, both of the new scalars are charged under the hidden neutrino-specific 
\Up which considerably alters phenomenology with respect to earlier works \cite{Gabriel:2006ns, Baek:2016wml, Bertuzzo:2017sbj, Nomura:2017wxf, Baek:2018wuo, Dey:2018yht}. 
Masses of the physical scalars, to leading order in $\xi\equiv\vbar/v_h$, are~\footnote{%
We use tadpole conditions to trade $\mu_{H,\Phi,S}$ for other parameters.
$m^2_{\Phi^\pm},m^2_{h_S}>0$ imply constraints on the parameter space, see \Figref{fig:mugamma}.}
\begin{align}
 m^2_H &= 2\,\lambda_H\,\vH^2\;, \quad m^2_{\Phi} = m^2_A  =  \frac{2\,\vH\,\mu}{s_{2\gamma}}\;,& \\
 m^2_{\Phi^\pm} &= \frac{\vH\,\mu}{t_\gamma}-\frac{\lambda_4}{2}\vH^2\;, & \\
 m^2_{\hS} &\approx \xi^2 \vH^2 \left(2\lambda_S-\frac{\lambda_{HS}^2}{2\lambda_H}\right) +\mathcal{O}(\gamma\mu/\vH)\;.&
 \end{align}%
The neutral scalar mass matrix is diagonalized by three orthogonal rotations $O=R(\theta_{13})R(\theta_{12})R(\theta_{23})$, such that
\begin{equation}
 O^\mathrm{T}\,M^2_{\mathrm{n.s.}}\,O=\mathrm{diag}\left(m^2_{\hS},m^2_H, m^2_{\Phi}\right)\;.
\end{equation}
The mixing angles, to leading order in $\xi$, are given by
\begin{align}
 s_{12}\equiv s_{S\Phi} &= s_\gamma\;, \quad s_{13}\equiv s_{HS} = \xi\,\frac{p\,t_\gamma  + q}{2\,\vH\,\lambda_H}\;, &\\
 s_{23}\equiv s_{\Phi H} &= \xi\,s_\gamma\,\frac{\mu \left(p\,t_\gamma  + q\right) - 2\lambda_H\,\vH\,p}{2\,\lambda_H\,\vH \left(\lambda_H\,\vH\,s_{2\gamma}-\mu\right)}\;,&
 \end{align}
where we use $\lambda_{34}:=\lambda_3+\lambda_4$ and
\begin{align}
 p:=\lambda_{34}\,v_H\,s_\gamma - \mu\,c_\gamma\;, \quad q:= \lambda_{HS}\,v_H\,c_\gamma-\mu\,s_\gamma.
\end{align}%
For the parameter region envisaged to resolve the Hubble tension, 
there are two new light bosonic fields: next to $Z'$ there is a 
scalar $h_S$ with mass in the $\keV$ range.

To prevent possible reservations about these light states straightaway, let us discuss their coupling to the SM.
The only way in which $h_S$ couples to fermions other than neutrinos is via 
mixing with the SM Higgs. Operators involving $h_S$ linearly can be written as 
$\mathcal{O}_{h_S}=c_{S\Phi}s_{HS}\times\mathcal{O}^{\mathrm{SM}}_{H\rightarrow h_S}$.
Hence, couplings to fermions are suppressed by their Yukawa couplings
and there are no new flavor changing effects. 
Ref.~\cite{Beacham:2019nyx} collects bounds on this scenario.
Besides BBN, discussed below, the strongest constraint arises from SN1987A burst duration and requires
$(c_{S\Phi}s_{HS})^2\lesssim10^{-12}$. Parametrically, $(c_{S\Phi}s_{HS})^2\sim\xi^2\sim\eM^4\times 10^{-10}$,
avoiding the bound for $\eM\lesssim0.1$.

The dominant coupling of $\Zp$ to SM fermions other than neutrinos is by $Z-\Zp$ mixing.
Given \eqref{eq:sinX}, $\Zp$ couples to the SM neutral current 
with strength $2g_X (\vP/\vH)^2=2g_X \xi^2 s_\gamma^2$. 
For momentum transfer below $\mZp$ this gives rise to new four-fermi (NSI) operators of effective strengths
\begin{subequations}\label{eq:NSI}
\begin{align}\label{eq:NSIa}
&\left(G^{(2\nu)(2f\neq\nu)}_{\mathrm{eff}}/G_{\mathrm{F}}\right)= -2\sqrt{2}\,\eM^2\,s_\gamma^2\;, \quad\text{and}&  \\\raisetag{16pt}
&\left(G^{(4f\neq\nu)}_{\mathrm{eff}}/G_{\mathrm{F}}\right)= 4\sqrt{2}\,\xi^2\,s_{\gamma}^4\approx\eM^4\,s_{\gamma}^4\,\times2\times10^{-9}.&
\end{align}
\end{subequations}
Such feeble effects are currently not constrained by experiment.

We note that the vector mixing of \eqref{eq:sinX} 
can be modified by gauge-kinetic mixing of the $\U1$ field strengths $\mathcal{L}_\chi=-(s_\chi/2)B^{\mu\nu}X_{\mu\nu}$~\cite{Galison:1983pa,Holdom:1985ag}.
This shifts the \Zp coupling to the SM neutral current by a negligible amount proportional to 
$\chi\,\mathcal{O}(\mZp^2/m_Z^2)$~\cite{Baumgart:2009tn, Babu:2017olk} (given $\mZp\ll m_Z,\chi\ll1$). 
More important is the introduction of a coupling of \Zp to the electromagnetic current 
scaling as $c_Wc_X\chi$. Experimental constraints on this are collected in \cite{Redondo:2015iea,Gherghetta:2019coi} 
and our model could, in principle, saturate these limits.
Therefore, we stress that $\chi\neq0$ would neither affect our solution to the HT, 
nor the $H$ and $Z$ decay rates in Eqs.~\eqref{eq:Gamhshs} 
and~\eqref{eq:GamZZp} below (to leading order),
which are fixed by Goldstone boson equivalence.

We thus focus on effects directly involving neutrinos.
For $T\lesssim v_\phi$, neutrino mixing is active. 
As required by direct-search bounds \cite{deGouvea:2015euy,Bryman:2019ssi,Bryman:2019bjg,Bolton:2019pcu}
and PMNS unitarity \cite{Antusch:2014woa} (as extracted from \cite{Fernandez-Martinez:2016lgt}) 
we assume~\cite{Farzan:2016wym,Farzan:2017xzy}
\begin{equation}\label{eq:epsilon-bounds}
 \eM^{(e)}\leq 0.050\;,\quad \eM^{(\mu)}\leq 0.021\;,\quad \eM^{(\tau)}\leq 0.075\;,
\end{equation}
for mixing with $e,\mu,\tau$ flavors. 
These are the most conservative bounds in the mass range set by Eq.~\eqref{eq:Mbound},
and often larger mixings can be allowed, especially for light masses and non-$e$ flavors.
For $M\lesssim10\eV$ oscillation experiments become important, requiring 
a dedicated analysis, see e.g.~\cite{Dentler:2018sju}, nontheless still allowing~\eqref{eq:epsilon-bounds}.
Since nothing in our resolution of the HT is required to depend on flavor, we adopt $\eM\sim5\times10^{-2}$ as a universal 
benchmark value.

Couplings of neutrinos to \Zp at low $T$ are given 
by $g_X\eM^2$ with a strength set by \eqref{eq:geffWindow}.
This gives rise to the four-fermion operators (\ref{eq:G4nueff},\ref{eq:NSI}),
but also to the possibility of \Zp emission off neutrinos.
Eq.~\eqref{eq:geffWindow} together with $g_X\lesssim1$ 
implies a \textit{lower} bound $\eM\gtrsim5\times10^{-4}$.
This fuels the intuition that this model is testable.

Constraints on neutrinos directly interacting with
light mediators are collected in \cite{Laha:2013xua, Ng:2014pca, Ioka:2014kca, Bustamante:2020mep,
Blinov:2019gcj,Escudero:2019gvw,Brdar:2020nbj,deGouvea:2019qaz,Lyu:2020lps}.
The strongest laboratory constraints arise from meson~\cite{Belotsky:2001fb, Lessa:2007up, Fiorini:2007zzc, Laha:2013xua, 
Bakhti:2017jhm,Dror:2017ehi,Dror:2017nsg,Bakhti:2018avv,Dror:2018wfl,Bahraminasr:2020ssz,Dror:2020fbh} 
and nuclear double-beta decays~\cite{Carone:1993jv,Agostini:2015nwa,Blum:2018ljv,Brune:2018sab,Deppisch:2020sqh}.
However, even the most stringent bounds for the least favorable flavor structure 
cannot exclude $g_{\mathrm{eff}}\lesssim10^{-5}$ for light $\mZp$.
While most experiments constrain light scalar (majoron) emission, 
the present study emphasizes the need to reanalyze those also for light vectors.
The most important constraint is SN1987A neutrino propagation through the cosmic neutrino background (C$\nu$B) \cite{Kolb:1987qy}.
The exact bound depends on the neutrino masses and rank of $y$, but even the most pessimistic assumptions 
allow $g_{\mathrm{eff}}\lesssim 5\times10^{-4}$ for $\mZp<60\,\eV$.

The $\bar{\nu}\nu\leftrightarrow\bar{\nu}\nu$ scattering cross section via $Z'$ exchange is approximated by
\begin{equation}
\sigma^{(4\nu)}(s)= \frac{g_X^4\,\eM^8}{12\,\pi }\frac{s}{\left(\mZp^2 - s\right)^2 + 
   \mZp^2 \Gamma_{\Zp}^2}\;. 
\end{equation}
To generate \Figref{fig:GamVsH} we include the $t$-channel and use Maxwell-Boltzmann 
thermal averaging \cite{Gondolo:1990dk},
while noting that a more refined analysis should employ  
Fermi-Dirac statistics \cite{Arcadi:2019oxh,Lebedev:2019ton}.
For $\mZp>2M_N$, $Z'$ decays to $\Nb1\N1$, $\Nb2\N2$, $\bar{\nu} \N2(\nu \Nb2)$ 
and $\bar{\nu} \nu$, while for $\mZp\lesssim M_N$ only the last channel is accessible.
The respective total widths are $\Gamma_{\Zp}/\mZp\approx 10^{-7}$ or $10^{-12}$,
corresponding to $Z'$ lifetimes from micro- to tens of picoseconds.
We show thermally averaged rates (dashed) obtained by dimensional analysis for $T\gg\vP$.
For $T\ll\mZp$ we reproduce the scaling of the effective operator \eqref{eq:G4nu}.
Obviously, before recombination $\Gamma_{4\nu}(T)$ differs from the effective theory.
This would not qualitatively change the conclusions of~\cite{Cyr-Racine:2013jua,Archidiacono:2013dua,Lancaster:2017ksf,Oldengott:2017fhy,Kreisch:2019yzn,Park:2019ibn},
which are based on the (non-)free-streaming nature of neutrinos, see also~\cite{Blinov:2020hmc}. 
Nontheless, this motivates a dedicated cosmological analysis to tell if our 
specific temperature dependence could be discriminated from the effective model.

Finally, we discuss the coupling of neutrinos to $\hS$.
For massless active neutrinos and conserved lepton number, 
diagonalization of \eqref{eq:Mnu} is exact in $s_{\eM}$.
This prevents a quadratic coupling of neutrinos to $\hS$.
Hence, SM neutrinos couple to $\hS$ only in association with hidden neutrinos, or 
suppressed by their tiny mass (e.g.\ $\left[M_\nu\right]_{11}\sim m_\nu$ produces such a coupling).
In both cases effects are unobservably small,
also because of the vastly suppressed coupling of $\hS$ to matter targets.

Also modification of $Z$ decays to neutrinos are unobservably small. 
Even if $N$ mixes with $\nu$, the invisible $Z$ width is unaltered for 
$M_N\ll m_Z$ \cite{deGouvea:2015euy}. The vertex $Z\bar{N}\nu$ also leads to 
$N$ production from neutrino upscattering on matter targets,
suppressed by $\eM$ compared to $G_{\mathrm{F}}$. 
$N$ decays invisibly leaving an unaccompanied recoil as signature.

Any consistent model of strong neutrino self-interactions 
requires a modification of the SM scalar sector. 
These are amongst the most visible effects of this model.
The necessary modifications imply new invisible decays of 
SM $Z$ and Higgs. To leading order in $\xi$, rates of the most prominent decays are
\begin{align}\raisetag{20pt}\label{eq:Gamhshs}
 \Gamma_{H\rightarrow\hS\hS} &= \frac{\vH^2}{32\,\pi\,m_H}\left[\lambda_{HS}\,c^2_\gamma+\lambda_{34}\,s^2_\gamma-\frac{\mu\,s_{2\gamma}}{\vH}\right]^2,& \\\raisetag{20pt}\label{eq:GamZZp}
 \Gamma_{H\rightarrow\Zp\Zp} &= \Gamma_{H\rightarrow\hS\hS}\;,\quad \Gamma_{Z\rightarrow\Zp\hS} = \frac{m_Z\,g_2^2\,s_{\gamma}^4}{192\,\pi\,c_W^2}\;,&
\end{align}%
and to leading order in $\xi$ and $\gamma$
(the exact expression in $\gamma$ is known)
 \begin{equation}
\Gamma_{H\rightarrow Z\Zp} \stackrel{\gamma\ll1}{=} 
\frac{g_2^2}{c_W^2}\frac{\left(m_H^2-m_Z^2\right)^3}{m_H^3\,m_Z^2}\frac{\,\xi^2\,s_\gamma^4\,}{16\,\pi}
\left(1+  \frac{\lambda_{HS}}{4\,\lambda_H}\right)^2\;. 
 \end{equation}
$\Gamma_{H\to\mathrm{inv.}}\leq1.3\,\MeV$ \cite{Tanabashi:2018oca,Cepeda:2019klc,ATLAS:2020cjb} 
and $\Gamma^{\mathrm{new}}_{Z\to\mathrm{inv.}}\leq2.0\,\MeV$ \cite{Carena:2003aj}
constrain the parameters as shown in \Figref{fig:mugamma}.
$\Gamma_{Z\rightarrow\Zp\hS}$ requires $\gamma\lesssim0.4$.
$\Gamma_{H\to\mathrm{inv.}}$, without fine tuning, demands $\lambda_{HS},\lambda_{34},\left(\mu s_{2\gamma}/\vH\right)\lesssim\mathcal{O}(10^{-2})$.
$\Gamma_{H\rightarrow Z\Zp}$ then is merely a rare Higgs decay with $\mathrm{BR}(H\rightarrow Z\Zp)\approx 10^{-8}\eM^4s_\gamma^4$.

The model without $S$ is excluded by $\Gamma_{Z\rightarrow\Zp\hS}$,
which would be given by \eqref{eq:GamZZp} with $\gamma\rightarrow\pi/4$.
The constraints on $\gamma$ further limit the allowed mass range of $M$, cf.~Eq.~\eqref{eq:Mbound},
and we show corresponding values of $M$ in~\Figref{fig:mugamma}.

Charged scalars $\Phi^\pm$ couple to $\ell^\pm$ and $N$ via \eqref{eq:Lint}. 
Constraints on $y$ arise from $\ell_1\to\ell_2\gamma$ and measured lepton magnetic moments, 
both mediated by a loop of $\Phi^\pm$ and $N$. 
Exact constraints are given in \cite{Farzan:2016wym}. 
Here it suffices to note that certainly $y\lesssim\mathcal{O}(1)$ for all flavors, 
as much tighter constraints are found below.

The coupling of $\Phi^\pm$ to quarks is suppressed by $s_\gamma\xi$ and 
standard LHC searches \cite{Aaboud:2018cwk,Sirunyan:2019arl} do not apply.
At LEP, $\Phi^\pm$ could have been pair-produced via $s$-channel $\gamma/Z$ 
or $t$-channel $N$, or singly-produced associated with charged and neutral leptons.
$\Phi^\pm$ dominantly decays to $N_\alpha\bar{\ell}_\beta$ with BRs set by $y$. 
$N$ further decays to neutrinos via $Z'$. The final state for $\Phi^\pm$ hence 
is $\ell_\beta^{\pm}+\mathrm{MET}$. LEP limits on $\Phi^\pm$ pair-production \cite{Abbiendi:2013hk} 
and a reinterpreted LEP selectron search~\cite{Heister:2001nk,Abbiendi:2003ji,Achard:2003ge,Abdallah:2003xe} 
imply a lower bound $m_{\Phi^\pm}> 100\GeV$~\footnote{%
Sometimes $m_{\Phi^\pm}\gtrsim275\GeV$ is quoted based on \cite{Khachatryan:2014qwa}. 
However, this requires assumptions on $\Phi^\pm$-BRs see \cite[Sec.~4.7]{Babu:2019mfe} for details.}.

Regarding electroweak precision, there are no new tree-level contributions to \mbox{$\rho\equiv\alpha\,T$}. 
We follow \cite{Grimus:2007if} for one-loop corrections. $T$ is always enhanced compared to the SM, staying
in the allowed interval $T=0.09\pm0.13$ \cite{Baak:2014ora} for $|m_{\Phi^\pm}-m_{\Phi}|\lesssim 120\,\GeV$.

\medskip
%\clubpenalty10000\widowpenalty10000
{\textbf{\textit{BBN.---}}}%
The bound on $\dNeff^{\mathrm{BBN}}$ during BBN \cite{Fields:2019pfx} 
prohibits any of the new light species, $Z'$, $N$, or $h_S$,
to be in thermal equilibrium with the SM during BBN.
Ultimately, a thermal QFT analysis seems worthwhile to fully explore the 
early universe cosmology of this model for $T\gg\vP$. 
The coupled Boltzmann equations then should be solved to track abundances precisely, 
but this is beyond the scope of this letter. 
Nontheless, an order of magnitude estimate suffices to clarify that there are 
parameters for which BBN proceeds as usual.

While $m_\Zp$ is fixed by \eqref{eq:mZp}, $M_N$ is limited by \eqref{eq:Mbound}, and 
$m_{\hS}\approx \xi \vH \sqrt{2 \lambda_S}\approx 7\eM^2\sqrt{\lambda_S}\,\MeV$.
Hence, neither of these states can simply be pushed beyond the $\MeV$ scale to avoid BBN constraints.
Instead, we discuss the possibility that all of them are sufficiently weakly coupled to the SM 
such that a thermal abundance is not retained.
While all of the new fields thermalize with the SM 
at EW temperatures, this changes once the heavy scalars freeze out.
The abundance of new light states is subsequently diluted by reheating 
in the SM. We thus focus on temperatures around BBN.

The coupling of $\Zp$ to the SM, as well as active-hidden neutrino mixing $\eM$,
is only effective after the \Up breaking phase transition \cite{Vecchi:2016lty}.
This warrants that $\Zp$ does not thermalize with the SM between EW and BBN.
Regarding $h_S$, the most relevant processes for thermalization are $e^+e^-\leftrightarrow h_S h_S$,
$e^-\gamma\leftrightarrow e^- h_S$, and $\nu\bar{\nu}\leftrightarrow h_S h_S$. 
None of them reaches thermal equilibrium due to the highly suppressed couplings of $h_S$.
The leading process thermalizing $N$'s with the SM is $e^+e^-(\nu\bar{\nu})\leftrightarrow N\bar{N}$ 
via $t$-channel $\Phi^\pm$ ($\Phi, A$) exchange, which scales as $\Gamma\sim (y/m_{\Phi^{(\pm)}})^4T^5$. 
Requiring this to be absent after QCD (EW) epoque demands \mbox{$y\lesssim 6\times 10^{-3(5)} (m_{\Phi^{(\pm)}}/100\,\GeV)$}.
Together with above bounds on $\eM $ and $\gamma\lesssim0.2$  this implies $M\lesssim 300(3)\eV$.
Consistency of our analysis requires $m_\nu\ll y\vP\ll M$, implying $y\gg 2\times10^{-5}(m_\nu/0.05\eV)$
and a lower bound on $\gamma$, shown in orange in~\Figref{fig:mugamma} for $m_\nu=0.05\eV$~\footnote{%
This is the minimal possible value of $m_\nu$ for at least one generation, as allowed by neutrino oscillation experiments.}.
Noteworthy, this limits $M_N$ to values which can resolve short baseline neutrino anomalies:
\textit{Either} in the well-known way with $\eV$-scale states, see e.g.~\cite{Dentler:2018sju, Boser:2019rta},
which however is already in considerable tension with other experiments~\cite{Dentler:2018sju}
and certainly requires specific assumptions on the flavor dependence of the mixing angle~\eM.
\textit{Or}, by realizing the novel ``decaying sterile neutrino solution'' for \mbox{$M_N\sim\mathcal{O}(100)\eV$}~\cite{Dentler:2019dhz,deGouvea:2019qre},
which requires the assumption $\eM^{e}\gg\eM^{\mu}$ \cite{Dentler:2019dhz}.

Finally, we note that despite bearing some danger for BBN, the new states can also be a virtue:
To resolve the HT with self-interacting neutrinos, \Neff must be enhanced to \mbox{$\dNeff^{\mathrm{CMB}}\approx1$} during 
recombination~\cite{Kreisch:2019yzn, Park:2019ibn}, requiring some energy injection in the dark sector after BBN.
The thermalization of $N, h_S, \Zp$ with $n$ flavors of active neutrinos after BBN produces entropy \cite{Hansen:2017rxr} 
which is released back to the neutrino background after the new states decay, before recombination. 
This would heat the neutrinos to $T_\nu'\approx[1+30/7n]^{1/12}T_\nu$
resulting in $\dNeff^{\mathrm{CMB}}\approx 1.03, 0.93, 0.74$ for $n=3,2,1$.
Regarding the decays of the light states to neutrinos, we note that 
$h_S$ decays to $\Nb1 \nu$ and $\Nb1\N2$, but in absence of $L$-violation not to 
$\bar{\nu}\N2$, $\Nb{1(2)}\N{1(2)}$ or $\bar{\nu}\nu$ (or all processes barred),
with $\Gamma_{h_S}\propto y^2 s_\gamma^2$. $\tau_{h_S}$, therefore, 
is extremely dependent on the exact parameters ranging from milli- to picoseconds.
Two-body decays $N\to Z'\nu$ contribute, provided $M_N>\mZp+m_\nu$. For $M_N<\mZp+m_\nu$, on the other hand, 
only three-body decays $N\to(2\nu)\bar{\nu}$ are possible with lifetime $\tau_N\sim(8\pi)^{-3}M_N^5{\Geff}^2\eM^{-2}$. 
Depending on the exact parameters, a population of $N$, thus, \textit{could} 
but but doesn't have to decay before recombination.

\medskip\widowpenalty10000
{\textbf{\textit{Discussion.---}}}%
We have presented a consistent (renormalizable and phenomenologically viable) 
model that leads to vector-mediated neutrino self-interactions. 
In a narrow region of parameter space these interactions have the right strength to 
resolve the tensions between local and global determinations of $H_0$ and $\sigma_8$~\footnote{%
Our model is viable also in a different parameter space not discussed here: 
For $\mZp\ll\eV$ and $g_X\ll 10^{-7}$, active neutrinos recouple only after 
recombination, with crucial impacts on the C$\nu$B.}.
To consistently implement such interactions in the SM, we had to introduce a second Higgs doublet and a new 
Dirac fermion, both charged under a new \Up gauge symmetry. Phenomenological consistency 
required the introduction of the \Up charged SM singlet scalar $S$.

There are several new states, all with very lepton-specific couplings: 
$h_S$ with mass of $\mathcal{O}(10\keV)$, as well as $\Phi$, pseudo-scalar $A$ and the charged scalars $\Phi^\pm$
all with masses of $\mathcal{O}(100\GeV)$.
The new, naturally neutrinophilic fore carrier has a mass of $\mZp\sim\mathcal{O}(10\eV)$, and the new hidden neutrinos 
masses in the range $M_N\sim 1\div300\eV$ and mixing angles $\eM>5\times10^{-4}$.
The allowed parameter space can be narrowed down by more precise measurements 
of $\mathrm{Higgs}\to\mathrm{inv.}$ and searches for leptophilic charged Higgses at the HL-LHC
and future colliders such as ILC, CLIC, or FCC-he/hh.
Other testable signatures include non-standard neutrino matter interactions with 
maximal strength $G^{(2\nu)(2f\neq\nu)}_{\mathrm{eff}}\sim\mathcal{O}(10^{-4})G_{\mathrm{F}}$,
as well as distortions of short baseline neutrino oscillations.

That our model works without specifying the mechanism of neutrino mass generation may feel 
like a drawback to some. We think this is a virtue, as it renders this scenario 
compatible with all standard neutrino mass generation mechanisms.

The least appealing feature of our model, perhaps, 
is the introduction of several new scales $(\vP,\vS,\mu,M)$, 
and some hierarchies among them. 
We have nothing to say here about this or any other hierarchy problem
but simply accepted this fact for the reason that we are convinced that this is 
the simplest renormalizable model in which \textit{active} neutrinos pick up gauged self-interactions.
Stabilizing these hierarchies against radiative corrections might require 
smaller scalar quartic cross-couplings than the direct constraints discussed above.
Suchlike would not contradict any of our findings.

Finally, our analysis also shows that ``model independent'' considerations are actually not always valid in concrete models.
On the contrary, only complete models allow to directly relate 
early universe cosmology, like the Hubble tension, to physics testable in laboratories.

\begin{acknowledgments}\vspace{-0.17in}
We would like to thank V.~Brdar, L.~Graf, and M.~E.~Krauss for useful conversations
as well as O.~Fischer, A.~Ghoshal, M.~Ratz, and S.~Vogl for comments on the manuscript.
In particular we thank S.~Vogl for pointing out the entropy production in the 
hidden sector re- and de-coupling dynamics.
This work benefited from using the computer codes \texttt{SARAH}~\cite{Staub:2013tta}, 
\texttt{FeynArts/FormCalc}~\cite{Hahn:2000kx,Hahn:1998yk}, and \texttt{PackageX}~\cite{Patel:2015tea}.

\medskip
\textbf{\textit{Note added.---}}%
During the completion of this work, Ref.~\cite{He:2020zns}
appeared on the arXiv which discusses a different 
UV complete model for self-interacting neutrinos.
\end{acknowledgments}

\bibliographystyle{utphys}
\bibliography{Orbifold}

\end{document}